\documentclass[runningheads,a4paper]{llncs}
\usepackage{amssymb} 
\usepackage{graphicx}
\usepackage{url}
\usepackage{float}
\usepackage{soul,color}
\usepackage{amsmath}
\usepackage{todonotes}
\usepackage{multirow}
\usepackage{longtable}
\usepackage[inline]{trackchanges}
\authorrunning{ }

\begin{document}

\title{Teaching DevOps in academia and industry: reflections and vision}

\author{Evgeny Bobrov \inst{1}, Antonio Bucchiarone \inst{3}, Alfredo Capozucca \inst{2} \\ Nicolas Guelfi \inst{2},  Manuel Mazzara \inst{1}, Sergey Masyagin \inst{1}
\institute{Innopolis University, Russian Federation \\
\and University of Luxembourg \\
\and Fondazione Bruno Kessler, Trento, Italy
%bucchiarone@fbk.eu
}
}

\toctitle{Lecture Notes in Computer Science}
\tocauthor{Authors' Instructions}
\maketitle
\begin{abstract}
This paper describes our experience of delivery educational programs in academia and in industry on DevOps, compare the two approaches and sum-up the lessons learnt. We also propose a vision to implement a shift in the Software Engineering Higher Education curricula.
\end{abstract}

\section{Introduction}
\label{sec:intro}

DevOps is a natural evolution of the Agile approaches \cite{DevOpsHandbook,Bass} from the software itself to the overall infrastructure and operations. This evolution was made possible by the spread of cloud-based technologies and the everything-as-a-service approaches. Adopting DevOps is however more complex than adopting Agile \cite{AgileDevops} since changes at organisation level are required. Furthermore, a complete new skill set has to be developed in the teams \cite{BucenaK17}. The educational process is therefore of major importance for students, developers and managers.

DevOps way of working has introduced a set of software engineering activities and corresponding supporting tools that has disrupted the way individual developers and teams produce software. This has led both the world of research and industry to review software engineering  life-cycle  and all the supporting techniques to develop software in continuous operation and evolution. If we want to enclose DevOps in one word, it is \textit{continuous}. Modelling, integration, testing, and delivery are significant part of DevOps life-cycle that, respect to enterprise or monolithic applications developed some years ago, must be revised continuously to permit the continuous evolution of the software and especially an easy adaptability at context changes and new requirements. Adopting the DevOps paradigm helps software teams to release applications faster and with more quality. In this paper, we consider two sides of the same coin that are the usage of DevOps in academia and in industry.
%Academic

Research in traditional software engineering settings has mainly focused on providing
batch automation, as in the case of translation and re-engineering of legacy code \cite{TrudelFNM13}, or on helping developers keep track of their changes, as in the case of version control \cite{EstlerNFM13}. The radically new development processes, introducing with the DevOps, have required major changes to traditional software practices \cite{DevOpsBook}. New versions of software components are developed, released, and deployed continuously to meet new requirements and fix problems.
%Industry
A study performed by Puppet Labs in 2015\footnote{https://puppet.com/resources/whitepaper/2015-state-devops-report} testifies that using DevOps practices and automated deployment led organisations to ship code 30 times faster, complete deployments 8,000 times faster, have 50\% fewer failed deployments, and restore service 12 times faster than their peers. Due to the dramatically growing of the DevOps supporting tools\footnote{\url{https://raygun.com/blog/best-devops-tools/}}, has seen a big change in the role played by the \textit{software engineers} of a team. The latter today have the complication of covering both management and development aspects of a software product. They are part of a team and have the following responsibilities: (1) to be aligned with the new technologies to ensure that the the high-performance software is released using smart tools to specify, develop, deploy and execute scalable software systems, (2) to define procedures to guarantee the high security level of the running code, (3) to monitor the software in operation and guarantee the right level of adaptability.

As long as DevOps became a widespread philosophy, the necessity of education in the field become more and more important, both from the technical and organisational point of view \cite{BucenaK17}. This paper describes parallel experiences of teaching both undergraduate and graduate students at the university, and junior professional developers in industry. There are similarities and differences in these two activities, and each side can learn from the other. We will discuss here some common issues and some common solutions. We also propose a vision to implement a shift in the Software Engineering Higher Education curricula.

The paper is organised as follows: after this introduction of the context in Section \ref{sec:intro}, we first discuss the experience gained in teaching DevOps at the university (Section \ref{sec:academia}). We then present the key elements of training and consultancies delivered in industry on the same subject (Section \ref{sec:industry}) and we analyse similarities and differences in Section \ref{sec:dicussion}. Section \ref{sec:vision} proposes a vision to implement a shift in the Software Engineering Higher Education curricula. Finally, in Section \ref{sec:conclusions} we present our conclusion.

\section{Teaching in Academia}
\label{sec:academia}

DevOps experienced significant success in the industrial sector, but still received minor attention in higher education. One of the few and very first courses in Europe focusing on DevOps was delivered at the university of Luxembourg \cite{CapozuccaGR18}.

%In this section we will discuss the evolution of the course in order to extract salient points and to compare with corporate education. In Section \ref{sec:vision} we will introduce a vision to redesign under a common frame Computer Science curricula for higher education.

%\subsection*{Context, Particularities and Targeted students}
This course is part of a graduate programme aimed at students pursuing a degree in computer science. Students following this programme either continue their development either in the private sector or doing a PhD at the same university (most of the cases). Therefore, most of the courses in such a programme are designed as a sequence of theoretical lectures and assessed by a mid-term and final exam. Our course is the exception in the programme as it is designed according to the Problem-based learning (PBL) method.

\subsection*{Organisation and delivery}
Following a problem-based approach, the learning of the students is centred on a complex problem which does not have a single correct answer. The complex problem addressed by the course corresponds to the implementation of a Deployment Pipeline, which needs to satisfy certain functional and non-functional requirements. These requirements are:

\begin{itemize}
\item Functional Requirements (FR)

\begin{itemize}
\item Create separated environments (Integration, Test, and Production)
\item Make use of a version control system
\item Make use of a continuous integration (CI) server
\item Automate the build of the selected product
\item Automate the execution of the test cases
\item Automate the deploy and release of the selected product
\end{itemize}

\item Non-Functional Requirements (NFR)

\begin{itemize}
\item Rely on technologies open-source and available for Unix-based OS 
\item The Product to test the functioning of the pipeline should be a Web App (SaaS) done in Java, if possible with an already available set of test cases
\end{itemize}
\end{itemize}

This means that students work in groups all along the course duration to produce a solution to the given problem. By working in groups students are immerse in a context where interactions problems may arise, and so allowing them to learn soft-skills to deal with such as problems. Therefore, the success to achieve a solution to the problem depends on not only the technical abilities, but also the soft-skills capacities each group member either has already had or is able to acquire during the course. Notice that DevOps is not only about tools, but also people and processes. Thus, soft-skills capabilities are a must for future software engineers working expected to work in a DevOps-oriented organisations.

\subsection*{Structure}
The course is organised as a mix of lectures, project follow-up sessions (aimed at having a close monitoring of the work done for each group member and helping solve any encountered impediments), and checkpoints (sessions where each group presents the advances regarding the project’s objectives). Lectures are aimed at presenting the fundamental DevOps-related concepts required to implement a Deployment Pipeline (Configuration Management, Build Management, Test Management, and Deployment Management). Obviously, the course opens with a general introduction to DevOps and a (both procedural and architectural) description of what a Deployment Pipeline is. In the first project follow-up session each group presents the chosen product they will use to demonstrate the functioning of the pipeline. The remaining of the course is an interleaving between lectures and follow-up sessions. The first check-point takes place at the fifth week, and the second one at the tenth week. The final checkpoint, where each group has to make a demo of the Deployment pipeline, takes place at the last session of the course. 

%Hands-on and content
\subsection*{Execution}
Most of the work done by the students to develop the Deployment Pipeline was done outside of the course hours due to the limited in-class time assigned to the course. However, examples (e.g. virtual environments creation, initial setup and provisioning) and references to well-documented tools (e.g. Vagrant, Ansible, GitLab, Jenkins, Maven, Katalon) provided during the sessions helped students on moving the project ahead. Moreover, students had to possibility to request support either upon appointment or simply signalling the faced issues with enough time in advance to be handled during a follow-up session. the teaching. Nevertheless, the staff was closely supervising the deployment pipeline development by both monitoring the activity on the groups’ working repositories and either asking technical questions or requesting live demos during the in-class sessions.

\subsection*{Assessment}
As described in \cite{CapozuccaGR18}, each kind of activity is precisely specified, so it lets students know exactly what they have to do. This also applies to the course assessment: while the project counts for 50\% of the final grade, the other half is composed of a report (12.5\%) and the average of the checkpoints (12.5\%). The aim at requesting to each group submit a report is to let students face with the challenge of doing collaborative writing in the same way most researchers do nowadays. Moreover, this activity makes the course to remain aligned with programme’s objectives: prepare the student to continue a research career. It is also in this direction the we have introduced peer-reviewing: each student is requested to review (at least one) no-authored report (this activity also contributes to the individual grading of the student). Despite of these writing and reviewing activities may seem specific to the programme where the course fits, we do believe that they also contribute to the development of the required skills software engineers need to have.

\subsection*{Latest experience and feedback}
Based on our latest experience the relevant points to highlight are: (1) the positive feedback obtained from students, (2) the absence of drops out, and (3) the quality of the achieved project deliverables. 
Regarding the first point, the evidence was found through a survey filled out by students once the course was over: 100\% strongly agreed that the course was well organised and ran smoothly, 75\% (25\%) agreed (strongly agreed) the technologies used in the course were interesting, and 75\% was satisfied with the quality of this course. 
We are very happy about the second point as it was one of the objectives (i.e. reduced the number of drops out - it used to reach up to 70\%) when we decided to redesign the course to its current format. Moreover, the absence of drops out can also be confirmed by the fact that (based on the survey) 75\% of the students would advise other students to take the course, if it were optional. Last, but not least, the survey also helped to confirm that PBL is the right pedagogical approach to tackle subjects like DevOps (and any others related to software engineering): 100\% of the students agreed that they would like to have more project-oriented courses like this one. 
The third relevant point was about the quality of the project deliverables: considering the limited time to present and work out the subjects related to a Deployment Pipeline, each group succeed to provide deliverables able to meet the given functional and non-functional requirements. 

\section{Teaching in Industry}
\label{sec:industry}

Our team is specialised in delivering corporate training for management and developers and has long experience of research in the service-oriented area \cite{Mazzara:phd,YanMCU07,YanCZM07}. In recent years we have provided courses, training and consultancies to a number of companies with particular focus on east Europe \cite{MazzaraNSSU18}. For example, only in 2018 more than 400 hours of training were conducted involving more than 500 employees in 4 international companies. Although we cannot share the details of the companies involved, they are mid to large size and employ more than 10k people.

The trainings are typically focusing on:

\begin{itemize}
\item Agile methods and their application \cite{AgileDevops}
\item DevOps philosophy, approach and tools \cite{Jabbari:2016}
\item Microservices \cite{Dragoni2017,DragoniLLMMS17}
\end{itemize}

\subsection*{Organisation and delivery}
In order for the companies to absorb the DevOps philosophy and practice, our action has to focus on people and processes as much as on tools. The target group is generally a team (or multiple teams) of developers, testers and often mid-management. We also suggest companies to include representatives from businesses and technical analysts, marketing and security departments. These participants could also benefit from participation and from  the DevOps culture. The nature of the delivery depends on the target group: sessions for management focus more on effective team building and establishment of processes. When the audience is a technical team, the focus goes more on tools and effective collaboration within and across the teams.

\subsection*{Structure}
The events are typically organised in several sessions run over a one-day to three-day format made or frontal presentations and practical sessions. The sessions are generally conducted at the office of the customer in a space suitably arranged after the previous discussion with the local management. Whenever possible the agenda and schedule of the activities have to be shared in advance. In this way, the participants know what to expect, and sometime a preparatory work is required.

\subsection*{Limitations of the set-up}
One of the limitations we had to cope with, often but not always, is the fact that bilateral previous communication with teams is not always possible or facilitated, and the information goes through some local contact and line manager. At times this demands for an on-the-fly on-site adaptation of the agenda. In order to collect as much information as possible on the participants and the environment, we typically send a survey to be completed a few days in advance, and we analyse question by question to give specific advice depending on the answers.

\subsection*{Lessons learnt and optimisation}
In retrospective, the most effective training for DevOps and Agile were those in which the audience consisted of both management and developers. Indeed the biggest challenge our customer encountered was not how to automatise existing processes, but in fact how to set up the DevOps approach itself from scratch. Generally, technical people know how to set up automatisation, but they may have partial understanding about the importance and the benefits for the company, for other departments, the customer and ultimately for themselves. It is important therefore to show the bigger picture and help them understanding how their work affects other groups, and how this in turn affects themselves in a feedback loop. The presence of management is very useful in this process. The technical perspective is often left for self-study or for additional sessions.

% Problem solving (to be melted in previous paragraph?)
%We eliminate the issue of not knowing the background of the participants in advance and even ask to create motley groups. We also propose to the companies include representatives from businesses and technical analysts, marketing departments and security departments due to the fact these people also get profit form DevOps cluture[ref].

\subsection*{Latest experience and feedback}
The feedback from participants surpassed our expectation. In synthesis, this are the major achievements of the past sessions:

\begin{itemize}
\item Marketers now understand how they may use A/B testing and check the hypothesis
\item Security engineers find positive to approve small pieces of new features, not the major releases
\item  Developers developed ways to communicate with other departments and fulfil their needs step by step based on the collaboration
\item Testers shifted their focus on product testing (integration-, regression-, soak-, mutation-, penetration- testing) rather than unit testing, and usually set future goals for continuing self-education on the subject
\end{itemize}

Often multiple session can be useful. The primary objective is to educate DevOps ambassadors, but it is also important to create an environment that can support the establishment of DevOps processes and the realisation of a solid DevOps culture, when every department welcome these changes. This does not typically happen in a few days.

\section{Discussion}
\label{sec:dicussion}

The experience of teaching in both an academic and industrial context emphasised some similarities and some differences that we would like to discuss here. Understanding these two realities may help in offering better pedagogical programme from the future since each domain can be cross-fertilised by the ideas taken by the other.

What we have seen in terms of similarities:

\begin{itemize}
\item \textbf{Pragmatism}: Both students and developers appreciate hands-on sessions
\item \textbf{Hype}: Interest and curiosity in the topic has been seen both in academia and industry, demonstrating the relevance of the topic
\item \textbf{Asymmetry}: Classic education and developers training put more important on Development than Operations and presenting the two sides as interrelated strengthen the knowledge and increase efficacy
\end{itemize}

What we have seen in terms of differences:
\begin{itemize}
\item \textbf{Learners initial state}: based on the academic curriculum where the course is included, it is possible to know (or at least to presume) the already acquired knowledge for the participant students. This may not be the case in a corporate environment, where the audience is generally composed by people with different profiles and backgrounds.
\item \textbf{Learners attitude}: students too often are grade-focused, developers are interested in the approach as long as can improve their working conditions, manager see things in terms of cost saving
\item \textbf{Pace of education}: short and intense in a corporate environment, can be long and diluted in academia
\item \textbf{Assessment/measure of success}: classic exam-based at the university, a corporate environment often does not require a direct assessment at the end of the sessions and the success should be observed in the long run
\item \textbf{Expectation:} corporate audience is more demanding. This may nor be a surprise given the costs and what is a stake. Students are also subject to a cost, but it is more moderate and spread over a number of course attended in one year.
\end{itemize}

\section{Vision}
\label{sec:vision}

After reporting experiences in teaching DevOps-based courses in both academic and industrial environments (reflection), in this section we will look at the future and we will describe our vision for the modernisation of university curricula in Computer Science, in particular for the Software Engineering tracks. While our vision and conclusions can be effectively applied in every Higher Education institution, we are here considering a specific case study: Innopolis University, a new IT educational institution in the Russian Federation. This is the reality we have more direct experience of. In \cite{Karapetyan18} the first five years of Innopolis University and the development of the internationalisation strategy is discussed, while \cite{Carvalho18} presents some teaching innovations and peculiarities of the university. At Innopolis University students have a 4-year bachelor, the firs two years are fundamental, and a specific track is chosen at the third year (Software Engineering, Data Science, Security and Network Engineering or Artificial Intelligence and Robotics). There are also 2-year Master Programs, following exactly the same four tracks. The last two years of the bachelors are characterised by a fewer number of courses. Moreover, some of these courses are elective, and delivered either by academic or industrial lecturers. These elective courses are aimed at covering specific topics required by industry. 

While working with industry we realised that the obstacles for the full adoption of DevOps are not only of technical nature, but also of mindset. This issue is difficult to solve since companies need to establish a radically new culture and transfer it to the new employees who join the company with a legacy mindset. The same situation may occur for fresh graduates. Classic curricula are very often based on the idea of \textit{system as a monolith and process as a waterfall}. Of course, in the last twenty years, innovations have been added to the plan of study worldwide. However, when focusing on the first two years of Bachelor education, it can be seen that the backbone of the curricula is still outdated (due to legacy reasons, and sometimes, ideological ones). It is therefore necessary to explain students the DevOps values from scratch, establishing clear connections of every course with DevOps, and describing how fundamental knowledge works within the frame of this philosophy. Furthermore, Computer Science curricula have a strong emphasis on the "Dev" part, but cover the "Ops" part only marginally, for example as little modules inside courses such as Operating Systems and Databases.

To cover the "Ops" part we need to teach how to engineer innovative software systems that can react to changes and new needs properly, without compromising the effectiveness of the system and without imposing cumbersome a priori analyses. To this end, we need to introduce courses on \textit{learning and adaptation} theories, algorithms and tools, since they are becoming the key enablers for conceiving and operating quality software systems that can automatically evolve to cope with errors, changes in the environment, and new functionalities. At the same time, to continuously assess the evolved system, we need also to think to teach \textit{validation} and \textit{verification} techniques pushing more them at runtime.

The DevOps philosophy is broad, inclusive, and at the same time, flexible enough to work as a skeleton for Software Engineering education. This is what drives our vision and we described in the next parts of this section.

\subsection{Phases of Software Engineering Education}

The DevOps philosophy presents recurring and neat phases. It has been shown that companies willing to establish a strong DevOps culture have to pay attention to every single phase \cite{DevOpsCulture}. Missing a phase, or even a simple aspect of it, might lead to poor overall results. This attention to every single phase should also be applied also to university education.

\begin{figure}[h]
\centering
\includegraphics[scale=0.6]{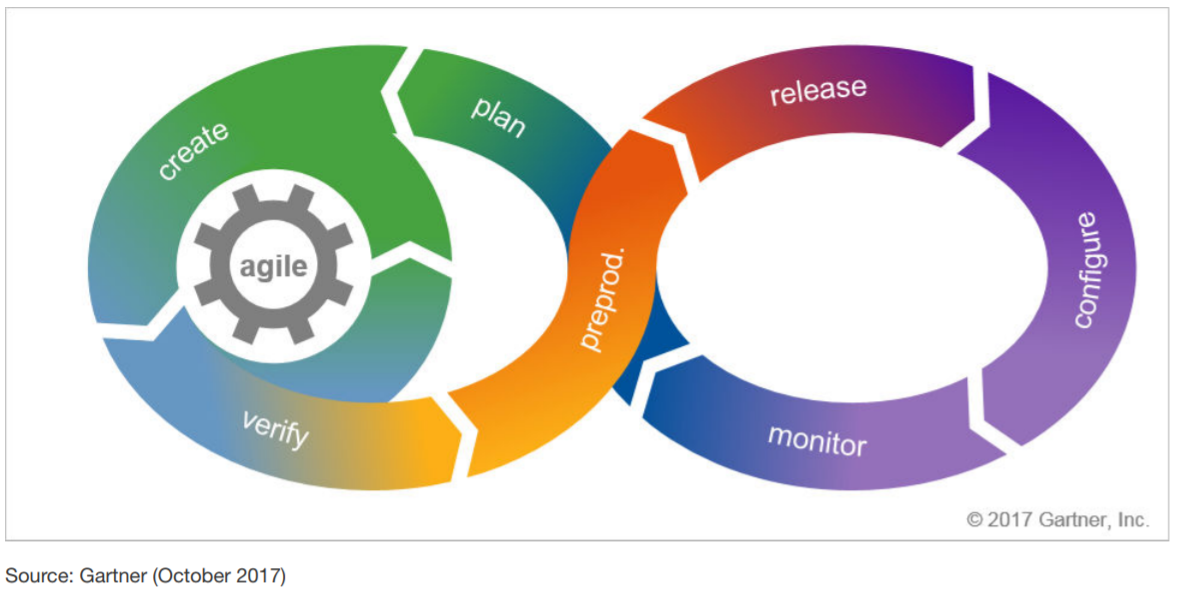}
\caption{DevOps Phases}
\label{fig:Phases}
\end{figure}

In this interpretation (or proposal), every phase corresponds to a series of concepts and a skill-set that the student has to acquire along the process. It is therefore possible to organise the educational process and define a curriculum for software engineering using the DevOps phases as a backbone (Figure \ref{fig:Phases} summarises these phases). This path would allow students to realise the connection between different courses and apply the knowledge in their future career. The plan described here is what we are considering to experiment at Innopolis University, expanding the experience acquired on the delivery of specialised DevOps courses to the entire plan of study. We will use the idea described in \cite{gartner_it_glossary_2017} as a backbone for curriculum innovation.

We consider ideal an incremental and iterative approach for bachelors to fully understand and implement the DevOps philosophy. We utilise the following taxonomy:

\begin{enumerate}
\item How to code
\item How to create software
\item How to create software in a team
\item How to create software in a team that someone needs
\item How to create software in a team that business needs
\end{enumerate}

In details, this is the path we propose for the bachelor\footnote{4-year} programme, based on Agile and DevOps according to the taxonomy:

\begin{enumerate}
\item The first three semesters are devoted to fundamental knowledge of hard and soft skills, which are essential to create software, especially following Agile and DevOps. We want to educate the next generations of students providing them not only with knowledge of programming languages and algorithms, but also with software architectures, design patterns and testing. This way students know how to create quality software fulfilling the essential non-functional requirements (such as reliability, maintainability, and scalability).

\item The fourth semester has a software project course (to be considered as an introduction to the software engineering track) based on the trial and error approach without any initial constraints and thorough analysis of identified problems in the second part of the semester.

\item The fifth semester has a new iteration of the software project course with a deep understanding of the Agile philosophy and the most popular Agile frameworks.

\item The sixth semester is based on the same project that has been created earlier and adds automaton, optimisation of the Development, and it introduces the Operational part and the feedback concept.

\item During the last two semesters (i.e. seventh and eighth), students start to work with real customers from industry and try to establish all processes and tools learnt in the previous three years.

\item During the third and fourth years, we propose additional core and elective courses in order to explore deeper modern technologies, best practices, patterns and frameworks.
\end{enumerate}

\subsection{Transition towards the new curriculum}
\begin{figure}[h]
\centering
\includegraphics[scale=0.363]{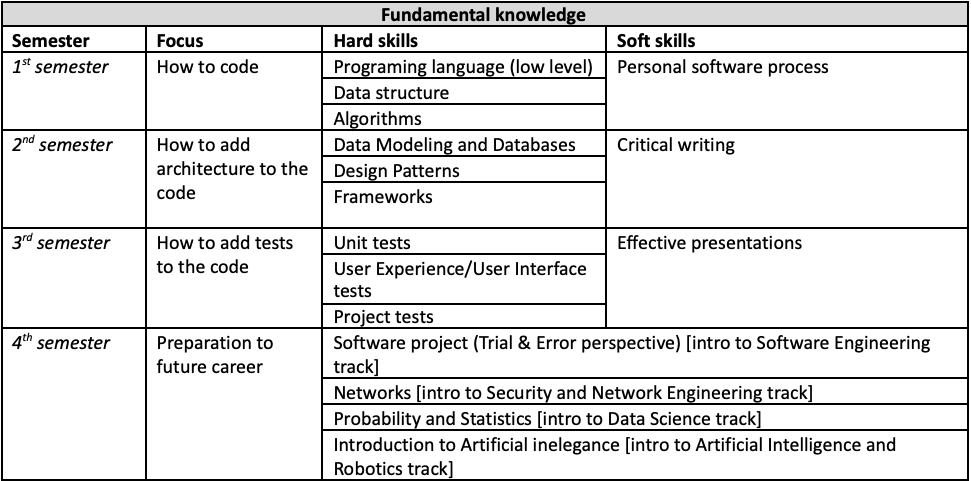}
\caption{Curriculum of Year 1 and Year 2}
\label{fig:BS1-2}
\end{figure}

\begin{figure}[h]
\centering
\includegraphics[scale=0.363]{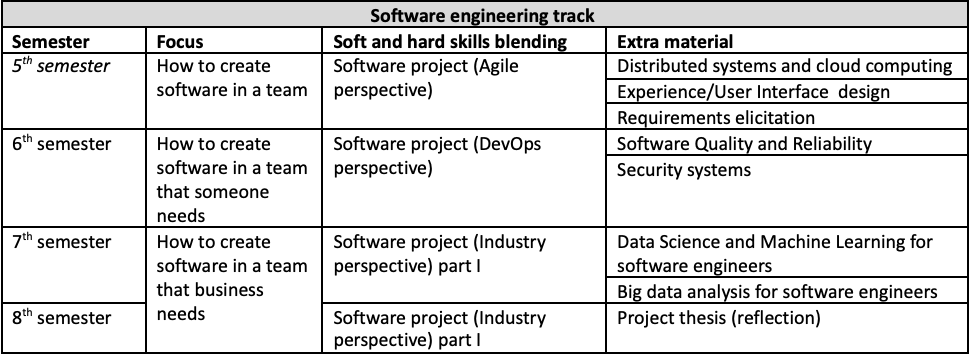}
\caption{Curriculum for Software Engineering track}
\label{fig:BS-SE}
\end{figure}

In this section we will address the transition from the current curriculum to the new one identifying the iterations and steps year by year until the full implementation, and we will emphasise the role of industry in this process. For the last years since foundation (2012), the curriculum for Software Engineers at Innopolis University was mostly waterfall-based with a clear focus on hard skills. Each course was delivering methods and tools specific of a certain  phase, but not always the \textit{"fil rouge"} between courses was emphasised. Courses connecting the dots and providing the basis for an iterative and incremental approach are now under development. The first four semesters of the bachelor provide the prerequisites for Software Engineering (and for Computer Science in general), whereas the last four semesters are track-based (see Fig.\ref{fig:BS1-2} and Fig.\ref{fig:BS-SE}).

The transition is planned to happen in 5-year time:

% [AC comment]: we need to add the semester when talking about a particular course
% [MM comment]: I added this detail wherever possible :-)

\begin{itemize}
\item \textbf{Year 1}. Make minor changes to the curriculum, targeting in particular two courses: \emph{Software Project} for second-year spring semester, and \emph{Project for Software Engineers} at the third year, fall semester. The first one has to be adapted to teach students how to establish processes and develop software according to Agile. The second one will be increased by adding the possibility to collaborate with industry and develop actual projects. The students interact with industry representatives and define project objectives with industry under the control of the university. 

% [AC comment]: the following statement is unclear. It has to be mentioned WHO reports to WHOM.
% The reports are with a two-week cadence from both students and industry. 
% [MM comment]: I agree that we can remove this sentence. It is not relevant here.

% [AC comment]: the strategy targets (or at least it was claimed to) address a change for the bachelor, thus no sense to talk about the Master programme.
% At the master level, a soft-skill course (\textit{"Communication"}) has been added with emphasis on public speaking, critical writing and business interaction.
% [MM comment]: correct - agree

\item \textbf{Year 2}. Work more closely with industry and add more elective courses covering skills required by companies. A course on DevOps will be added to the spring semester of the third year of the bachelor to be intended as a continuation of \emph{Software Project}. The content of some courses will be adjusted to contain DevOps philosophy.

\item \textbf{Year 3}. Update fundamental courses at the first and second year according to the Software Engineering Body of Knowledge (SWEBOK) standard \cite{IEEEComputerSociety:2014} (chapters “Mathematical Foundations”, “Computing Foundations” and “Engineering Foundations). Furthermore, soft skills courses such as \textit{“personal software process”}, \textit{“critical writing”} and \textit{“effective presentations”} will be added to the first three semesters.

\item \textbf{Year 4}. Follow the SWEBOK and deliver the most essential knowledge areas.

%from applying perspective for bachelors third and fourth courses and analyzing perspective for the master programme with respect to bloom taxonomy.

\item \textbf{Year 5}. Analyse the results of the changes introduced, and then tune the fundamental courses with more notions of DevOps and Agile philosophies along with incremental-iterative approaches. By year 5 we are planning to establish a framework helping to update the curriculum to give more focus on industry demands and IT evolution.
\end{itemize}

\section{Conclusions}
\label{sec:conclusions}

Ultimately, DevOps \cite{Bass,Jabbari:2016} and the microservices architectural style \cite{Dragoni2017,DragoniLLMMS17} with its domains of interests \cite{Salikhov2016a,Salikhov2016b,Nalin2016,BucchiaroneDDLM18,MazzaraTSC2018} may have the potential of changing how companies run their systems in the same way Agile has changed the way of developing software. The critical importance of such cultural change should not be undervalued. It is in this regard that higher education institutions should put a major effort to fine tune their curricula and cooperative programme in order to meet this challenge.

In terms of pedagogical innovation, the authors of this paper have experimented for long with novel approaches under different forms \cite{Carvalho18}. However, DevOps represents a newer and significant challenge. Despite of the fact current educational approaches in academia and industry show some similarities, they are indeed significantly different in terms of attitude of the learners, their expectation, delivery pace and measure of success. Similarities lay more on the perceived hype of the topic, its typical pragmatic and applicative nature, and the minor relevance that education classically reserves to "Operations". While similarities can help in defining a common content for the courses, the differences clearly suggest a completely different nature of the modalities of delivery. 

From the current experience we plan to adjust educational programs as follows:

\begin{itemize}
\item \textbf{University teaching}: trying to move the focus out of final grade, emphasising more the learning aspect and give less importance to the final exam, maybe increasing the relevance of practical assignments. It may be also useful to intensify the theoretical delivery to keep the attention higher and have more time for hand-on sessions. Ultimately, our vision is to build a Software Engineering curricula on the backbone derived from the DevOps philosophy.

\item \textbf{Corporate training}: it is important not to focus all the training activity as a frontal session university-like. Often the customers themselves require this classical format, maybe due to the influence of their university education. We believe that this makes things less effective and we advocate for a change of paradigm.
\end{itemize}

Finally, we have described our vision for the transition to the new curriculum at Innopolis University. In terms of educational innovation, other realities are moving fast and we should not be shy in proposing curricula drastic changes. Looking, for example, at domains like security and dependability, the Technical University of Denmark (DTU) is modernising the approach at both MSc and continuous education programme levels~\cite{SxCMatching2009,DRAGONI20091628}.

\bibliographystyle{unsrt}
\bibliography{biblio}

\end{document}